\documentclass{emulateapj}

\usepackage{xspace}
\usepackage{graphicx}
\usepackage{natbib}
\usepackage{apjfonts}
\newcommand\tabspace{\noalign{\vspace*{0.8mm}}}

\def\errtwo#1#2#3{$#1^{+#2}_{-#3}$}

\newcommand\zth{$0^\mathrm{th}$\xspace}
\newcommand\fst{$1^\mathrm{st}$\xspace}

\newcommand\acis{{ACIS}\xspace}
\newcommand\acisi{{ACIS-I}\xspace}
\newcommand\aciss{{ACIS-S}\xspace}

\newcommand\chandra{\textsl{Chandra}\xspace}

\newcommand\epic{{EPIC}\xspace}

\newcommand\heg{{HEG}\xspace}
\newcommand\hetg{{HETG}\xspace}
\newcommand\hetgs{{HETGS}\xspace}

\newcommand\meg{{MEG}\xspace}

\newcommand\rosat{\textsl{ROSAT}\xspace}

\newcommand\swift{\textsl{Swift}\xspace}

\newcommand\xmm{\textsl{XMM-Newton}\xspace}
\newcommand\rgs{{RGS}\xspace}

\newcommand\sgra{{Sgr A$^*$}\xspace}

\newcommand\aproxgt{\mathrel{%
     \rlap{\raise 0.511ex \hbox{$>$}}{\lower 0.511ex \hbox{$\sim$}}}}
\newcommand\aproxlt{\mathrel{%
     \rlap{\raise 0.511ex \hbox{$<$}}{\lower 0.511ex \hbox{$\sim$}}}}

\slugcomment{Submitted 2012 June 14.  Resubmitted 2012 September 4.  Accepted $\ldots$}

\shorttitle{The Brightest Flare Seen from \sgra} 
\shortauthors{Nowak et al.}

\begin{document}

\title{Chandra-HETGS Observations of the Brightest Flare Seen from
  \sgra} \author{M. A. Nowak\altaffilmark{1},
  J. Neilsen\altaffilmark{1}, S. B. Markoff\altaffilmark{2},
  F. K. Baganoff\altaffilmark{1}, D. Porquet\altaffilmark{3},
  N. Grosso\altaffilmark{3}, Y. Levin\altaffilmark{4,5},
  J. Houck\altaffilmark{1}, A. Eckart\altaffilmark{6},
  H. Falcke\altaffilmark{7,8,9}, L. Ji\altaffilmark{10},
  J. M. Miller\altaffilmark{11}, Q. D. Wang\altaffilmark{12}}
\altaffiltext{1}{Massachusetts Institute of Technology, Kavli
  Institute for Astrophysics, Cambridge, MA 02139, USA;
  mnowak,jneilsen,fkb,houck@space.mit.edu}
\altaffiltext{2}{Astronomical Institute, ``Anton Pannekoek'',
  University of Amsterdam, Amsterdam, The Netherlands;
  s.b.markoff@uva.nl} \altaffiltext{3}{Observatoire Astronomique de
  Strasbourg, Universit\'e de Strasbourg, CNRS, UMR 7550, 11 rue de
  l'Universit\'e, 67000 Strasbourg, France;
  delphine.porquet,nicolas.grosso@astro.unistra.fr}
\altaffiltext{4}{School of Physics and CSPA, Monash University, VIC
  3800, Australia; Yuri.Levin@monash.edu}
\altaffiltext{5}{Leiden Observatory, Leiden University, P.O. Box 9513,
  NL-2300 RA Leiden, The Netherlands}
\altaffiltext{6}{Physikalisches Institut, Universit\"at zu K\"oln,
  Z\"ulpicher Str. 77, 50937 K\"oln, Germany; eckart@ph1.uni-koeln.de}
\altaffiltext{7}{Department of Astrophysics, IMAPP, Radboud University
  Nijmegen, PO Box 9010, 6500 GL Nijmegen, The Netherlands;
  H.Falcke@astro.ru.nl}
\altaffiltext{8}{ASTRON, Postbus 2, 7990 AA Dwingeloo, the
  Netherlands}
\altaffiltext{9}{Max-Planck-Institut f\"ur Radioastronomie, Auf dem
  H\"ugel 69, 53121 Bonn, Germany}
\altaffiltext{10}{Purple Mountain Observatory, CAS, Nanjing,
  P.R. China, 210008; ji@pmo.ac.cn} 
\altaffiltext{11}{Department of Astronomy, University of Michigan, Ann
  Arbor, Michigan, MI 48109, USA; jonmm@umich.edu}
\altaffiltext{12}{Department of Astronomy, University of
  Massachusetts, Amherst, MA 01002, USA; wqd@astro.umass.edu}

\begin{abstract}
Starting in 2012, we began an unprecedented observational program
focused on the supermassive black hole in the center of our Galaxy,
\sgra, utilizing the {High Energy Transmission Gratings Spectrometer}
(\hetgs) instrument on the \textsl{Chandra X-ray Observatory}.  These
observations will allow us to measure the quiescent X-ray spectra of
\sgra for the first time at both high spatial and spectral resolution.
The X-ray emission of \sgra, however, is known to flare roughly daily
by factors of a few to ten times over quiescent emission levels, with
rarer flares extending to factors of greater than 100 times
quiescence.  Here were report an observation performed on 2012
February 9 wherein we detected what is the highest peak flux and
fluence flare ever observed from \sgra.  The flare, which lasted for
5.6\,ks and had a decidedly asymmetric profile with a faster decline
than rise, achieved a mean absorbed 2--8\,keV flux of
$(8.5\pm0.9)\times10^{-12}\,{\rm erg~cm^{-2}~s^{-1}}$. The peak flux
was 2.5 times higher, and the total 2--10\,keV emission of the event
was approximately $10^{39}$\,erg.  Only one other flare of comparable
magnitude, but shorter duration, has been observed in \sgra by \xmm in
2002 October.  We perform spectral fits of this \chandra observed
flare, and compare our results to the two brightest flares ever
observed with \xmm.  We find good agreement among the fitted spectral
slopes ($\Gamma\sim2$) and X-ray absorbing columns ($N_{\rm
  H}\sim15\times10^{22}$ cm$^{-2}$) for all three of these events,
resolving prior differences (which are most likely due to the combined
effects of pileup and spectral modeling) among \chandra and \xmm
observations of \sgra flares.  We also discuss fits to the quiescent
spectra of \sgra.
\end{abstract}


\section{Introduction}\label{sec:intro}

\setcounter{footnote}{0}

Sagittarius (Sgr) A* is the compact radio, infrared and X-ray source
associated with the $4\times10^6\;M_\odot$ supermassive black hole at
the dynamical center of our Galaxy (see, e.g.,
\citealt{MeliaFalcke2001,GenzelEisenhauerGillessen2010}).  As the
nearest galactic nucleus, \sgra offers unique access to accretion
physics on event horizon scales, and is thus a key testbed for
theoretical modeling.  On the other hand in the almost 40 years since
its radio identification \citep{BalickBrown1974}, a multitude of
observational campaigns in the radio/millimeter, near-infrared (NIR)
and X-ray bands have established that \sgra is emitting steadily at a
bolometric luminosity $L_{\rm Bol}\sim 10^{-9}\,L_{\rm Edd}$, orders
of magnitude lower than is typical for nearby low-luminosity active
galactic nuclei (LLAGN; e.g., \citealt{Ho1999}).  \sgra is thus either
representative of a distinct class of quiescent galactic nuclei
lurking within most normal galaxies, or it is simply occupying the
extreme low end of the AGN continuum (see, e.g., \citealt{nagar:05a},
Figure~4).  Placing \sgra into context with other objects is therefore
an important goal, in order to correctly interpret its rather atypical
features.

The X-ray band is a powerful probe of the inner accretion flow regions
of black holes.  The \textsl{Chandra X-ray Observatory} soon after its
launch was the first to identify \sgra by discovering a dominant,
steady emission state \citep{Baganoffetal2003} which can just be
spatially resolved at \chandra's subarcsecond imaging resolution.
This emission can be associated with thermal bremsstrahlung from near
the gravitational capture radius (\citealt{Quataert2002}; but see
\citealt{sazonov:12a} for an alternative).  The X-ray ``quiescent
state'' is punctuated roughly daily by flares with $\aproxlt$ hour
time scales that point to a source within $\sim10$'s of $r_g \equiv
GM/c^2$ (the gravitational radius) from the black hole
\citep{Baganoffetal2001,Goldwurmetal2003,Belangeretal2005,Porquetetal2003,Porquetetal2008}.
The flares, whose emission has been modeled with synchrotron or
alternatively inverse Compton, are most likely caused by nonthermal
processes \citep[e.g., shock or magnetic reconnection acceleration of
  electrons within a
  jet;][]{Markoffetal2001,YuanQuataertNarayan2004,Yusef-Zadehetal2006,Dodds-Edenetal2009},
though other mechanisms have also been suggested
\citep[e.g.,][]{LiuMelia2002,ZuvobasNayakshinMarkoff2012}.

Simultaneous monitoring campaigns with the NIR have established that
all the X-ray flares have NIR counterparts, while only the brighter
NIR flares ($\ge 10$ mJy) have corresponding X-ray activity
\citep{Eckartetal2004,Ghezetal2004,hornstein:07a,Dodds-Edenetal2009,Dodds-Edenetal2011,trap:11a}.
The relationship of X-ray and NIR flaring to that in the
sub-millimeter (submm) bands is still under debate, although broad
peaks delayed from the NIR have been noted
\citep{marrone:08a,Yusef-Zadehetal2006,yusefzadeh:08a}.

Many important questions persist about the nature of the accretion
flow in \sgra, and particularly about the flares, which seem to
provide a missing link with activity seen in other weakly accreting
black holes.  Specifically, in the last decade the Fundamental Plane
(FP) of black hole accretion
\citep{MerloniHeinzDiMatteo2003,FalckeKoerdingMarkoff2004,KoerdingFalckeCorbel2006}
has emerged as an important concept that links black hole mass and
radiative output in the radio and X-ray bands for weakly accreting
systems, with \sgra being an extreme low-luminosity example of such
systems.  As statistics have improved, it now appears that the X-ray
flux of \sgra approaches or meets the expectations from the FP only
during its flaring state, but lies at too low an X-ray flux, relative
to its radio emission, during quiescence
\citep{Markoff2005,KoerdingFalckeCorbel2006,Plotkinetal2012}.  As the
FP radio luminosity is associated with synchrotron emission from
accelerated particles in compact jets, the flares may be providing
clues about jet launching and plasma processes near the black hole.

Only about two dozen X-ray flares have been detected so far, primarily
with \chandra and \xmm. Most flare fluxes are on the order of a factor
of a few to ten times the quiescent flux, but a few show fluxes on the
order of 100 times greater than quiescence, and sometimes have
associated pre- or post-cursor ``hiccups'', i.e., weak flares close in
time to the major outburst
\citep{Baganoffetal2001,Porquetetal2003,Porquetetal2008}.  Exact flare
characteristics such as spectral slope, which is very important for
constraining models, are not well-determined because \chandra flares
suffer pileup.  Pileup in the Advanced CCD Imaging
Spectrometer-Imaging array (\acisi; \citealt{garmire:03a}) is when two
or more events arrive in overlapping pixel regions within the same
detector readout frame, and subsequently are either read as a single
event with the summed energy or are discarded as a non-X-ray event
\citep{davis:01a}.  Although \xmm flare observations do not suffer
from pileup, its larger mirror point spread function (PSF) does not
isolate the accretion flow as effectively. Prior studies had shown
spectral slope differences between flares observed with \chandra and
\xmm \citep{Baganoffetal2001,Porquetetal2003,Porquetetal2008}, and it
has been unclear to what extent these differences were due to
instrumental effects.

As part of an unprecedented X-ray Visionary Project (XVP) awarded in
its Cycle 13 Guest Observer Program, \chandra has begun the first of a
total of 3\,Msec of observations to be carried out in 2012 with the
{High Energy Transmission Grating Spectrometer} (\hetgs;
\citealt{canizares:05a}).  The main goal of this program is to
constrain the accretion processes around \sgra, including detecting
flares for the first time with high-resolution spectroscopy ($E/\Delta
E\approx 200$@6.4\,keV).  In our first several observations, already
two major flaring periods have occurred, the first of which contains
the brightest flare detected to date.  This flare is sufficiently
strong and long to allow us to create an individual spectrum.  Here we
present a detailed analysis of this flare and compare its spectra and
characteristics to the two brightest X-ray flares reported in the last
decade of \sgra observations.  In addition we discuss the
quiescent state, as observed in the \zth order of the gratings, and
suggest a standardized method for reporting flare peaks and spectra to
aid in characterizing the flare distribution for studies of the entire
sample of events.

\section{Observations}
\label{sec:obs}
At the time of writing of this paper, 2012 April 1, there have been 39
\chandra observations of \sgra using its \acisi, totaling 1.2\,Msec,
and now 10 using the \hetg in combination with the \acis-Spectroscopy
array (\aciss), totaling 320\,ks.  The \hetg is comprised of two
gratings sets: the medium energy gratings (\meg) and the high energy
gratings (\heg), which disperse spectra into positive and negative
spectral orders. We consider only the \heg and \meg $\pm$\fst orders,
which between them cover the $\approx 0.5$--9\,keV energy band.
Additionally, an on-axis undispersed image is created at CCD spectral
resolution (the \zth order). Compared to observations without
insertion of the \hetg, the \zth order efficiency ranges from $\approx
30\%$ at 2\,keV to $\approx 80\%$ at 8\,keV, with an average of
$\approx 40\%$ when weighted by \sgra's quiescent count rate spectrum.

The focus of our present work is a very bright flare observed with the
\hetgs during a 59.25\,ks observation that began at 06:17:03 UTC on
2012 February 9 (ObsID 14392). The overall reduction in \zth order
efficiency means that \sgra flares like this one are subject to
significantly less photon pileup than \acisi observations conducted
absent the gratings. (\aciss spectra also have slightly higher
spectral resolution than \acisi spectra with comparable S/N.) The
dispersed gratings spectra of \sgra flares are never subject to
pileup.  We leave the challenging analysis of the quiescent gratings
spectra, which include significant background from diffuse Galactic
center emission dispersed across the field of view, for a later date,
when more data are available and we have a more reliable model of the
background emission.

We took two steps to isolate \sgra's flare emission and minimize the
contribution from diffuse X-ray background.  First, we extracted
spectra and lightcurves from small circular regions with radii of 2.5
pixels ($\approx 1$\arcsec\hspace{-3 pt}.25) centered on \sgra's
celestial coordinates\footnote{We did not re-register coordinates,
  since the latest \chandra data processing versions register \sgra to
  a positional accuracy of typically 0\arcsec\hspace{-3 pt}.1.}. For
the major flare discussed in this paper, we also extracted lightcurves
and spectra for the $\pm$\fst order gratings (as determined by the
\texttt{tg\_resolve\_events} tool) from long, 5 pixel wide rectangular
regions centered on the dispersed spectra. Second, we reprocessed and
extracted CCD spectra from all existing \chandra \sgra observations,
including \acisi data, to provide the best possible characterization
of the quiescent spectrum.  All data were processed with standard
\texttt{CIAO v4.4} tools \citep{fruscione:06a} and calibration
database \texttt{v4.4.8}. We selected standard event grades (0, 2--4,
6) and applied corrections for Charge Transfer Inefficiency (CTI), but
did not apply pixel randomization to the event positions.  For \acisi
and \hetgs \zth order spectra, detector response matrices and
effective areas were created with the \texttt{mkacisrmf} and
\texttt{mkarf} tools\footnote{\acisi response matrices for ObsID 292,
  however, were created with the \texttt{mkrmf} tool as this
  observation occurred at a -110 C focal plane temperature.}, with the
effective areas being ``aperture corrected'' using the
\texttt{arfcorr} tool.  (This tool divides the effective area by an
energy-dependent fraction that ranged from $\approx 0.9$--0.83 between
2--8\,keV, which accounts for the fraction of the energy-dependent,
point source PSF within the 2.5 pixel radius source region.)  Gratings
responses were created with the \texttt{mkgrmf} (which includes flux
aperture correction) and \texttt{mkgarf} tools \citep[see][for an
  outline of the gratings processing procedures]{daveh:11a}.

\section{Lightcurves}

\begin{figure}
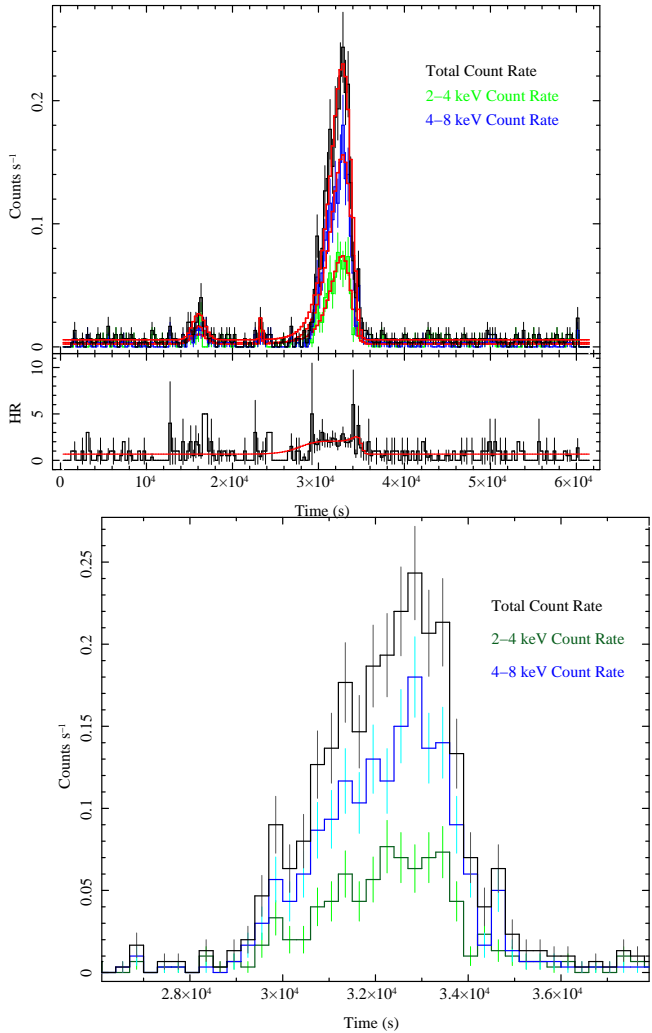

\begin{center}
\includegraphics[width=0.45\textwidth,viewport=20 20 565 495]{SgrA_Flare_HR}
\includegraphics[width=0.45\textwidth,viewport=20 20 565 495]{SgrA_Flare_zoom}
\end{center}
\caption{Top: 2--8\,keV, 2--4\,keV, and 4--8\,keV \sgra lightcurves in
  300\,s bins from \chandra ObsID 14392, comprised of \zth order and
  $\pm$\fst order counts. Time is measured relative to the observation
  start: 2012 Feb.\ 9, 06:17:04 UTC. Each lightcurve is fit with a
  constant, two Gaussian distributions (for the precursor flares), and
  a Gumbel distribution.  Middle: Hardness ratio of the
  4--8\,keV/2--4\,keV rates, shown with the hardness ratio from the
  fits.  Bottom: Close-up of the bright flare, highlighting its
  asymmetric profile.}
 \label{fig:flare}
\end{figure}

We searched for flares in the \sgra lightcurves by applying a Bayesian
Blocks algorithm (Scargle 2002, priv. comm.) to unbinned events in the
2--9\,keV band, using the implementation from the \texttt{S-lang/ISIS
  Timing Analysis
  Routines}\footnote{\texttt{http://space.mit.edu/cxc/analysis/SITAR/}}
(\texttt{SITAR}).  This same method was employed previously by
\citet{Baganoffetal2003}.  We chose a detection significance level of
98.2\%, i.e., $1-\exp(-4)$, for each lightcurve ``change point''.
Since a flare has at least two change points, a rise and a decay, the
overall flare significance is at least 99.97\%. Each lightcurve is
then described by a series of uniform rate bins (usually only one bin
for the mean rate).  For lightcurves described by multiple bins, any
bin with a rate below the 2-$\sigma$ upper bound of the lowest rate
bin was considered as ``quiescent'', while the remaining bins were
assigned as ``flare''.  Contiguous flare bins were considered to be a
single flare, and were excised to create quiescent spectra for all
observations.  Using the Bayesian Blocks algorithm, we found 18 flares
in the \acisi observations and 6 flares in the \hetgs observations,
with 2 in ObsID 14392.  In a future work we will describe the
statistics and properties of the full set of flares detected in all
\chandra observations; however, here we are concerned with the second
flare from ObsID 14392.  Both the mean and peak count rates from this
flare were significantly higher than any other observed \chandra
flare.

Figure~\ref{fig:flare} presents the full, energy-resolved X-ray
lightcurve for ObsID 14392, comprised of \zth and $\pm$\fst order
events in 300\,s bins.  A large flare occurs roughly halfway through
the observation and lasts $\approx5$\,ks, with two possible precursor
flares at $\sim16,000$\,s and $\sim22,000$\,s. The latter flare
appears in the Bayesian Blocks lightcurve only if we decrease the
detection significance level (for detecting two change points) to
93\%. As discussed in Section~\ref{sec:spectra}, we create spectra for
the mean of the large flare, but do not have sufficient statistics to
describe the spectra at the flare peak.  Determining the amplitude of
the flare in the lightcurve is therefore particularly important for
assessing the flare's peak luminosity.

\begin{figure}
\begin{center}
\includegraphics[width=0.45\textwidth, viewport=0 5 555 380]{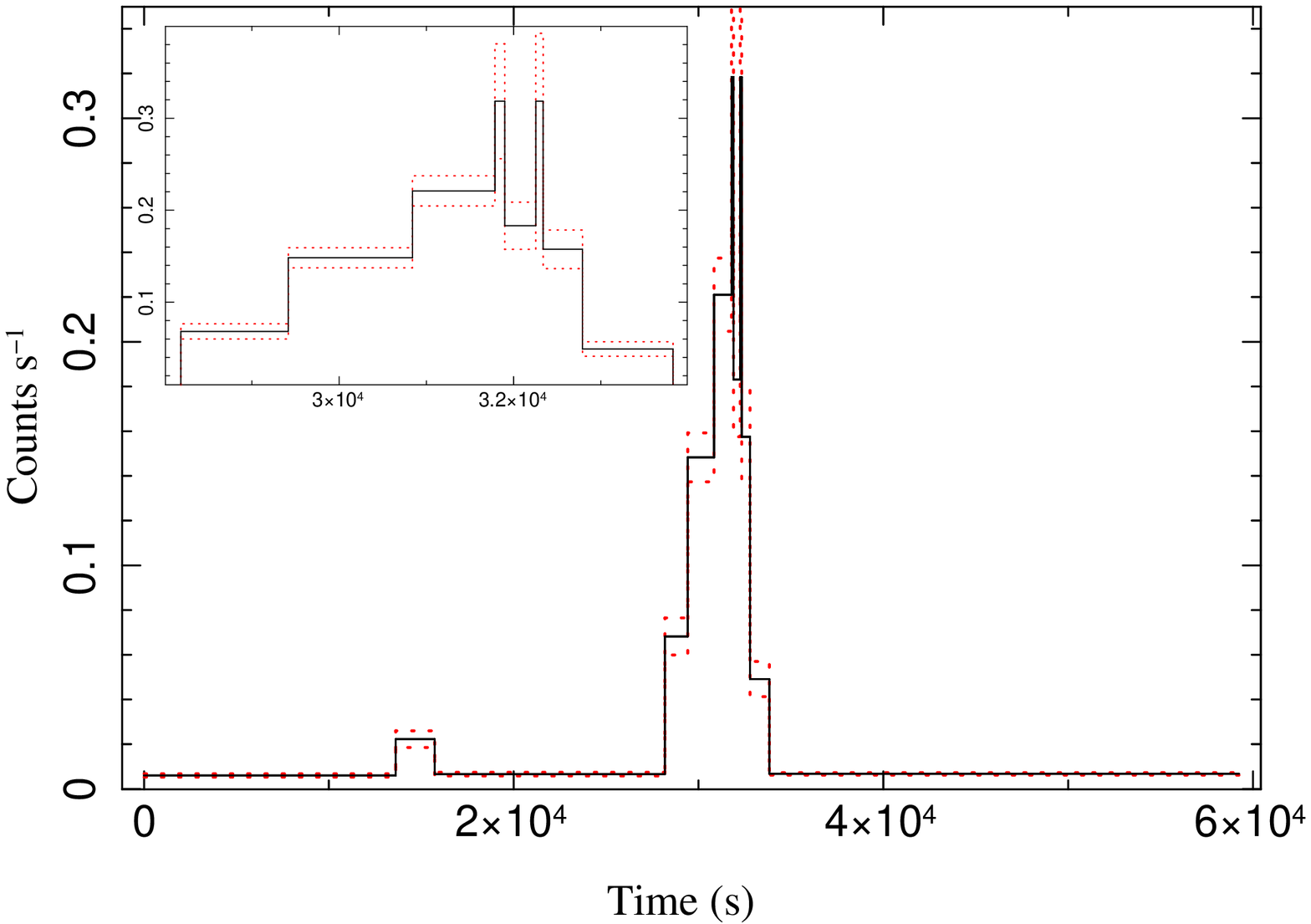}
\includegraphics[width=0.45\textwidth, viewport=80 20 580 385]{rate_corrII}
\end{center}
\caption{Top: For ObsID 14392, the summed \zth and \fst order count
  rates (solid line) with $1\sigma$ errors (dotted line) during
  intervals found by the Bayesian Blocks algorithm (using a 98.2\%
  significance level for each change point).  The major flare start
  and stop times are 2012 Feb. 9, 14:25:50 and 15:59:51 UTC.  Bottom:
  For each Bayesian Block, after subtracting the mean quiescent level
  rate, the ratio of the \zth order rate to the summed \fst order
  rates vs. the summed \fst order rates. (Error bars are 1$\sigma$.)
  The light blue line is the expected correlation if the intrinsic
  (i.e., unpiled) \zth order rate is 1.7$\times$ the \fst order rate,
  and the pileup parameter $\alpha=1$.}
 \label{fig:BB}
\end{figure}

The precursor flares are rather faint; however, the main flare is
clearly visible above the background level, which is composed of the
quiescent emission from \sgra as well as diffuse emission throughout
the extraction regions.  The main flare is asymmetric
(Figure~\ref{fig:flare}, bottom), with a slow rise and a sharp decline,
and can be modeled as a strong, wide ($\sigma\sim1400$s) Gaussian
flare followed by a weaker, narrower ($\sigma\sim400$s) Gaussian flare
approximately 1100 s later. However, since our primary focus is in
determining the peak brightness of the flare, it is useful to have
this quantity as a free parameter when fitting the lightcurve. We
accomplish this with a renormalized Gumbel distribution:
\begin{equation}
 f(t)=N_{\rm peak}~e^{(t-t_{0})/\tau}~
  e^{1-\exp[(t-t_{0})/\tau]} ~~,
\end{equation}
where $N_{\rm peak}$ is the peak count rate, $t$ is time, $t_0$ is the
peak time, and $\tau$ is the characteristic time scale. This provides
a good match to the flare's asymmetry (see the red curves in
Figure~\ref{fig:flare}). Our full model for the lightcurve consists of
Gaussians for the two precursor flares and a Gumbel component for the
main flare, superimposed on a constant baseline. The mean flare-only
count rate, calculated over the interval where the main flare is
brighter than the background level, is $0.091\pm0.006$
counts\,s$^{-1};$ the peak flare-only count rate is $0.22\pm0.2$
counts\,s$^{-1},$ and the characteristic time scale for the flare is
$1180\pm70$\,s. The error bars are 90\% confidence intervals for a
single parameter using the Cash statistic \citep{cash:79a}. Thus the
flare peak rate is $2.5\pm0.3$ times the mean flare count rate. This
should be regarded as a lower limit, since the \zth order lightcurve
is actually suppressed by pileup during the flare (see
Figure~\ref{fig:BB} and the discussion below).

There is evidence for substructure in our Bayesian blocks
decomposition of the flare (Figure~\ref{fig:BB}): the peak of the flare
is consistent with having a brief $\approx 300$\,s dip in between two
sharp peaks of $\approx 100$\,s duration. This structure exists
independently (albeit at lower significance) in both the \zth order
and summed \fst order lightcurves. A similar dip/short time scale
structure was seen in the 2002 October flare observed by \xmm
\citep{Porquetetal2003}.

\begin{figure}
\begin{center}
\includegraphics[width=0.38\textwidth, viewport=20 20 430 410]{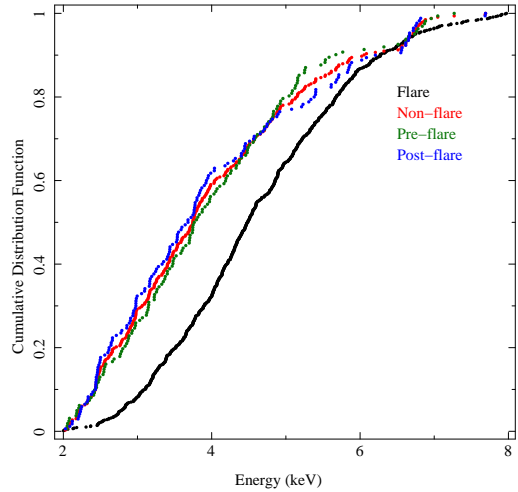}
\end{center}
\caption{Cumulative distribution functions (CDFs) of the \zth and
  $\pm$\fst order counts as a function of energy for the main flare
  and the quiescent interval, further subdivided into pre- and post-
  main flare intervals. The quiescent and flare CDFs differ at $>95\%$
  confidence. The rise in the quiescent CDF at 6.6 keV indicates a
  strong contribution from the known iron emission.}
 \label{fig:KS}
\end{figure}

We also have searched for any X-ray color differences between the
flare and non-flare intervals. First, we calculated a hardness ratio
(HR) as the ratio of the 4--8\,keV lightcurve to the 2--4\,keV
lightcurve. Zeros in the denominator were replaced with the median
2--4\,keV count rate. The results are shown in the bottom panel of
Figure~\ref{fig:flare}. The smooth red curve is the ratio of the fits to
the relevant light curves, and the flare appears harder
than the quiescent emission. A Kolmogorov-Smirnov (K-S) test on the
extracted events indicates that the probability that the flare and
quiescent intervals have the same spectrum is $P=3\times10^{-15}$ (see
Figure~\ref{fig:KS} for the cumulative distribution functions, i.e.,
CDFs). If we consider only events from the $\pm$\fst order gratings
spectra or the \zth order, the probability that the flare and
quiescent intervals have the same spectrum is $P=3\times10^{-9}$ and
$P=4.6\times10^{-2},$ respectively. We conclude that at the $>95\%$
level, the flare spectrum is harder than the quiescent spectrum. There
is no evidence for a difference in the pre-flare and post-flare
spectra, although the CDFs in Figure~\ref{fig:KS} do indicate that the
$\approx6.6$\,keV iron emission \citep{Baganoffetal2003,sazonov:12a}
is relatively significant in the quiescent spectrum, even in this
short observation.

\begin{deluxetable*}{lccccccccccc}
\setlength{\tabcolsep}{0.03in} 
\tabletypesize{\footnotesize}    
\tablewidth{0pt} 
\tablecaption{Joint Fit to Chandra Flare and Quiescent Spectra 
              from the \sgra Region\label{tab:flux}}
\tablehead{
     \colhead{State}
   & \colhead{$F^{\mathrm{abs}}_{2-8}$}
   & \colhead{$F^{\mathrm{unabs}}_{2-8}$}
   & \colhead{$F^{\mathrm{unabs}}_{2-10}$}
   & \colhead{$L^{\mathrm{unabs}}_{2-10}$}
   & \colhead{$F^{\mathrm{abs}}_{\nu}$}
   & $\mathrm{N}_{\mathrm{H}}$
   & $\Gamma$
   & \colhead{$E_{\mathrm{line}}$}
   & \colhead{$\sigma_{\mathrm{line}}$}
   & \colhead{EW}
   & $\chi^2$/DoF
          \\                               
   & \multicolumn{3}{c}{($10^{-12}~\mathrm{erg~cm^{-2}~s^{-1}}$)}
   & ($10^{34}~\mathrm{erg~s^{-1}}$) 
   & (nJy)
   & ($10^{22}~\mathrm{cm^{-2}}$) 
   &
   & \multicolumn{3}{c}{(keV)}
         }
\startdata
   Flare Mean
 & \errtwo{8.5}{0.9}{0.9}       
 & \errtwo{21.6}{10.3}{5.2}     
 & \errtwo{25.1}{9.4}{4.8}      
 & \errtwo{19.2}{7.2}{3.7}      
 & {770}                        
 & \errtwo{14.3}{4.4}{3.6}      
 & \errtwo{2.0}{0.7}{0.6}       
 & \nodata                      
 & \nodata                      
 & \nodata                      
 & 267/256                      
\\
\tabspace
   Quiescent
 & \errtwo{0.147}{0.004}{0.003} 
 & \errtwo{0.45}{0.04}{0.04}    
 & \errtwo{0.47}{0.05}{0.03}    
 & \errtwo{0.36}{0.04}{0.04}    
 & {9.6}                        
 & \errtwo{12.9}{0.8}{0.8}      
 & \errtwo{3.0}{0.2}{0.2}       
 & \errtwo{6.63}{0.02}{0.02}    
 & \errtwo{0.02}{0.04}{0.02}    
 & \errtwo{0.78}{0.14}{0.12}    
 & 267/256                      
\\
\tabspace
   Flare Peak
 & \errtwo{21}{3}{3}            
 & \errtwo{54}{27}{15}          
 & \errtwo{63}{25}{14}          
 & \errtwo{48}{19}{11}          
 & \nodata                      
 & \nodata                      
 & \nodata                      
 & \nodata                      
 & \nodata                      
 & \nodata                      
 & \nodata                      
\enddata 

\tablecomments{The model,
  \texttt{dustscat$\times$TBnew$\times$(powerlaw+gaussian)}, is
  applied separately to the quiescent spectrum and the flare
  spectrum. No gaussian is included during the flare. We model the
  \zth order spectrum during the flare as the sum of the quiescent
  emission and flare emission; for the \hetg flare spectrum, we
  subtract the quiescent emission and diffuse extended emission as
  background. In lieu of model normalizations, we measure the
  integrated 2--8 and 2--10 keV fluxes $F$ with the \texttt{cflux}
  convolution model. We report both absorbed (superscript abs) and
  unabsorbed (superscript unabs) fluxes; the 2--10\,keV luminosity $L$
  presumes isotropic emission at a distance of
  8\,kpc. $F^{\mathrm{abs}}_\nu$ is the best fit model flux density at
  6\,keV (no errors given). $\mathrm{N}_\mathrm{H}$ is the equivalent
  hydrogen column density, $\Gamma$ is the power law index, and
  $E_\mathrm{line},$ $\sigma_\mathrm{line},$ and EW are the energy,
  $1\sigma$ width, and equivalent width of the Gaussian emission line.
  Errors are 90\% confidence level for one interesting parameter.
  Peak flux values are derived assuming a peak/mean flux ratio of
  $2.5\pm0.3$ (90\% CL), with errors combined in quadrature. Due to
  pileup, this ratio may in fact be $\approx 10\%$ too low (see
  text).}
\end{deluxetable*}  

\begin{figure*}[ht]
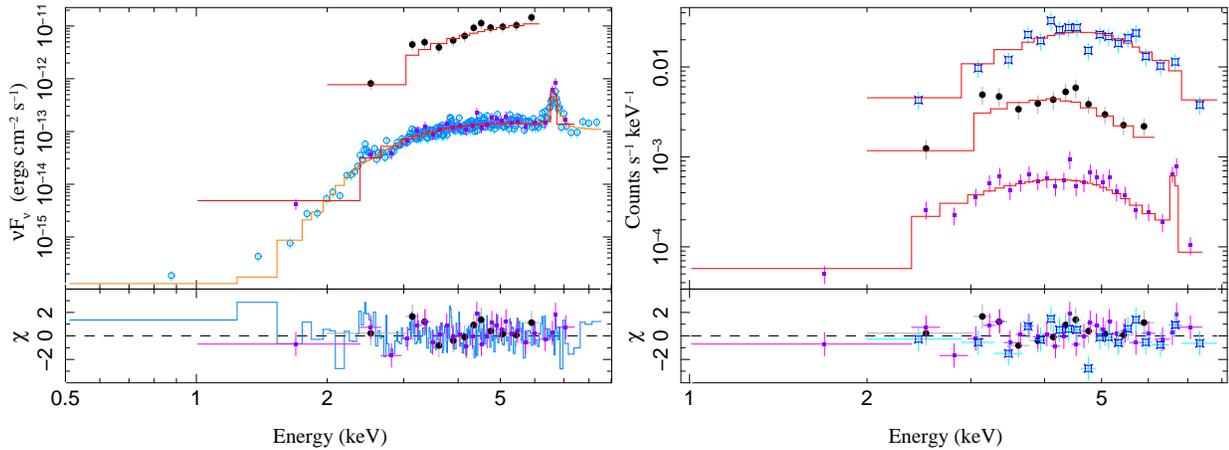

\begin{center}
\includegraphics[width=0.45\textwidth, viewport=85 20 579 380]{all_data_flux}
\includegraphics[width=0.45\textwidth, viewport=85 20 579 380]{all_data_density}
\end{center}
\caption{Left: Summed, flux corrected spectra for the inner
  1\arcsec\hspace{-3 pt}.25 radius surrounding \sgra, with the fit
  presented in Table~\protect{\ref{tab:flux}}.  Spectra represent:
  quiescent emission (\acisi, hollow blue circles; and \hetg \zth
  order, solid purple squares), and the bright flare mean emission
  (\hetg $\pm$\fst orders, solid black circles).  Right: The same
  spectra (absent \acisi spectra) shown as detector
  $\mathrm{counts~s^{-1}~keV^{-1}}$, and now including \zth order
  spectra for the flare (hollow blue squares).}
 \label{fig:spectra}
\end{figure*}

To search for evidence of any color evolution during the flare, we
examine the ratio of the \zth order to summed \fst order count rates,
as the former is most sensitive to hard X-rays while the latter is
most sensitive to soft X-rays.  Figure~\ref{fig:BB} shows this rate
ratio for the lightcurve sub-intervals obtained from the Bayesian
Blocks decomposition.  Because the \zth order is still subject to
pileup, this ratio can vary even in the absence of color
evolution. For a constant spectral shape, the ratio of the \zth order
counts and $\pm$\fst order counts should scale as:
\begin{equation}
  (\alpha\Lambda_i)^{-1} \left [ \exp \left ( \alpha \Lambda_i \right ) - 1 \right ] 
  \exp \left ( - \Lambda_i \right ) ~~,
\label{eq:pile}
\end{equation} 
where $\Lambda_i$ is the incident (unpiled) counts per integration
frame, and the odds of $N$ piled photons being detected as a single
``good event'' is assumed to be $\propto \alpha^{(N-1)}$
\citep{davis:01a}. The expected curve is shown in the bottom panel of
Figure \ref{fig:BB}, under the assumption that all piled events are
recorded (i.e., $\alpha=1$) and that the gratings rate is
$\approx$37\% of the total incident unpiled count rate. The data are
consistent with no detectable spectral evolution during the flare,
aside from pileup effects. We estimate that due to the effects of
pileup, our lightcurve measurement of the peak/mean ratio is a factor
$\approx 1.1$ too low.

These naive estimates, however, are subject to systematic
uncertainties in the details of the pileup model. For the \zth order
spectra, we estimate that on average, 8\% of the \emph{incident} flare
photons are subject to pileup, with this fraction increasing to 16\%
at the flare peak. Of the \emph{detected} \zth order flare events, on
average 0--4\% (for $\alpha=0$--1) are in fact ``piled'' events
falsely registered at higher energies that harden the spectrum.  This
systematic effect is accounted for in the \zth order spectral fits
below, aided by the fact that the simultaneously fit \hetgs spectra
are not subject to pileup.  Nevertheless, uncertainty in pileup
modeling (specifically, $\alpha$) serves to widen the error bars on
the photon index fit to the flare spectra.

\section{Spectra}\label{sec:spectra}

We next consider the spectrum for the bright flare observed in ObsID
14392, and specifically compare it to \sgra quiescent spectra.  It is
not completely straightforward to characterize \sgra's quiescent
emission, as it is clearly extended
\citep{shcerbakov:10a,sazonov:12a}.  In fact, recent models of the
extended emission \citep[i.e.,][]{shcerbakov:10a} suggest that only on
the order of 1\% of the observed quiescent emission arises near the
event horizon, as opposed to flare models where, owing to time scales
of only thousands of seconds, almost all the emission is associated
with the inner region.  We ignore these distinctions and do not break
up the quiescent emission into ``point-like'' and ``extended''
components, nor do we even attempt to ``background subtract'' the
quiescent emission.  Instead, we use consistent extraction regions
between the quiescent and flare periods, and treat the flare as
\emph{additional} emission. Since our quiescent spectra include
diffuse emission, we generally prefer to report absolute flare flux
levels, rather than describing the flare in terms of a ``factor times
quiescent emission.'' The mean absorbed flux density in the 2--8 keV
band is the quantity least subject to systematic uncertainty and most
useful in comparing current and prior observations.

\subsection{Methodology}
To accomplish these measurements, we created a \zth order spectrum and
$\pm$\fst order spectra for the 5600s interval of the brightest flare
in ObsID 14392. In order to isolate the actual flare spectrum, it is
important to have a reliable characterization of the quiescent
spectrum. For the gratings, we created a background spectrum by
extracting a \fst order spectrum from the quiescent periods of ObsID
14392. Because the \zth order spectral analysis of the flare is
complicated by the presence of pileup, we modeled the \zth order flare
spectrum as the sum of a flare component and a quiescent component. To
constrain the quiescent component, we created quiescent spectra from
\textit{all} \chandra observations of \sgra. We kept the \acisi and
\zth order \aciss spectra separate, but fit them with a single
model\footnote{All analyses described in this paper have been
  performed with the \texttt{Interactive Spectral Interpretation
    System} (\texttt{ISIS}; \citealt{houck:00a}).  Spectra for each
  ObsID and gratings arm were kept separate, but like groups were
  combined during analysis using the \texttt{combine\_datasets}
  function. Plots also show the combined data, and when showing ``flux
  corrected'' data this correction has been implemented using only the
  detector response matrices, and not the fitted models.}. This method
appears justified because our K-S test does not indicate a difference
between the pre- and post-flare quiescent spectra, and because prior
studies have suggested a fairly stable quiescent level
\citep{shcerbakov:10a}.

\begin{figure*}
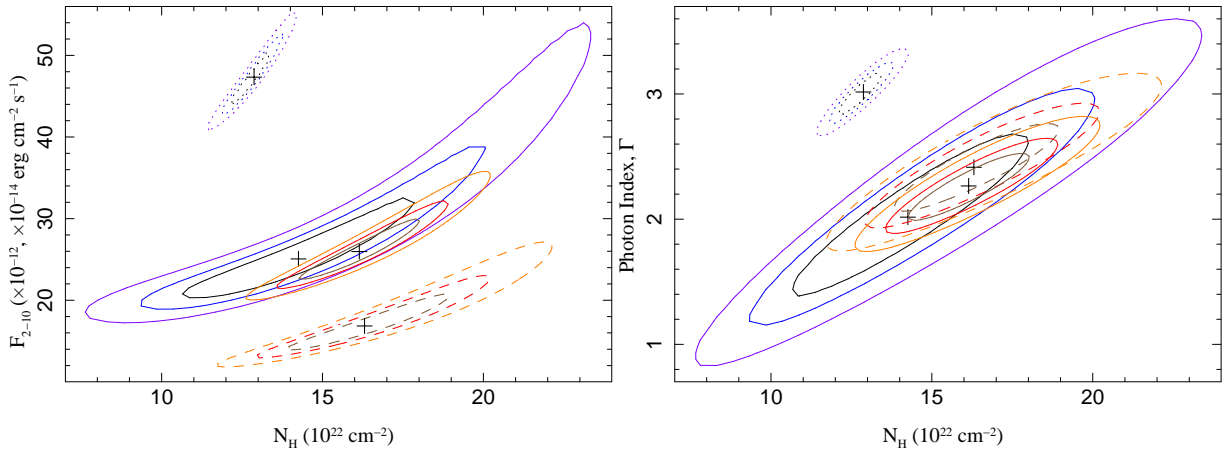

\begin{center}
\includegraphics[width=0.45\textwidth, viewport=85 20 579 380]{new_flux_nh_contour}
\includegraphics[width=0.45\textwidth, viewport=85 20 579 380]{new_gamma_nh_contour}
\end{center}
\caption{Confidence contours for \sgra spectral parameters at several
  epochs: mean unabsorbed 2--10\,keV flux vs. X-ray absorbing column,
  ${\rm N_H}$, (left) and photon index vs. ${\rm N_H}$ (right).
  Black/blue/purple lines are for the quiescent (dotted lines) and
  major flare (solid lines) emission observed by \chandra in ObsID
  14392.  Brown/red/orange lines are for flare emission observed by
  \xmm on 2007 April 4 (dashed lines) and on 2002 October 3 (solid
  lines) \protect{\citep{Porquetetal2003,Porquetetal2008}},
  re-analyzed with the same spectral model as applied to the \chandra
  observation.  Contours are 68\%, 90\%, and 99\% significance for two
  interesting parameters.  Flare flux is in units of $10^{-12}\,{\rm
    erg\,cm^{-2}\,s^{-1}}$ and quiescent flux is in units of
  $10^{-14}\,{\rm erg\,cm^{-2}\,s^{-1}}$.}
 \label{fig:conf}
\end{figure*}

We restricted our fits of the quiescent \acisi spectra to the 0.5--9
keV band, the \zth order quiescent spectra to the 1--9 keV band, and
the gratings spectra to the 2--9 keV band. All combined spectra were
rebinned to have a minimum signal-to-noise of 4.5 in each energy bin,
and the gratings spectra were further required to have a minimum of 16
pre-binning channels per final energy bin. Only those bins completely
inside the above energy ranges were included in our analysis.

For clarity, we set out the details of our fitting process. The
quiescent spectra from \acisi and the \zth order were fit jointly with
a single model consisting of an absorbed, dust-scattered power-law and
an iron emission line (see below for details). The \fst order grating
spectrum of the flare was modeled as a second absorbed, dust-scattered
power-law, and the \zth order spectrum of the flare was treated as the
sum of these two components convolved through the {\tt ISIS} pileup
model. (The pileup parameter $\alpha$ was left as a free parameter,
but its error bars always spanned the full range 0--1. Again, this
uncertainty contributes to widening the error bars on the fitted
photon index, $\Gamma$.)

Although the overall spectral model is rather simple, the interstellar
absorption and dust scattering components merit further
discussion. X-ray absorption is dominated by metals, not hydrogen,
hence the fitted hydrogen column is strongly dependent on the adopted
cross sections and abundances \citep{wilms:00a}. We use the
\texttt{TBnew}
model\footnote{http://pulsar.sternwarte.uni-erlangen.de/wilms/research/tbabs/index.html}
developed from the work and abundances described by
\citet{wilms:00a}. In our experience, using {\tt TBnew} with the cross
sections of \citet{verner:96a} yields equivalent ${\rm N_H}$ values
$\sim 1.5\times$ that of the oft used \texttt{wabs} model
\citep{morrison:83a}.

For all intents and purposes, given the very small \chandra PSF, dust
scattering acts as a pure loss term, with the dust scattering optical
depth having an $E^{-2}$ dependence. We use the model
\texttt{dustscat} \citep[see][]{Baganoffetal2003}, which has an
optical depth at 1\,keV proportional to the X-ray absorbing column
density. This proportionality has been measured via dust scattering
halo images and X-ray binary spectra obtained with \rosat by
\citet{predehl:95a}, who found $\tau \approx 0.486\,({\rm
  N_H}/10^{22}\,{\rm cm^{-2}}$) when using an analog of the
\texttt{wabs} model and cross sections from \citet{morrison:83a}.
Given the rough scaling between {\tt TBnew} and {\tt wabs}, we tie our
dust scattering optical depth to our fitted equivalent ${\rm N_H}$ via
$\tau=0.324\,({\rm N_H}/10^{22}\,{\rm cm}^{-2}).$ The implied
extinction (from the correlations of \citealt{predehl:95a}) is then
${\rm A_V} \sim {\rm N_H^{TBnew}}/2.69\times10^{21}\,{\rm
  cm^{-2}}$. However, these dependencies have not been revisited with
modern absorption or dust scattering models (e.g.,
\citealt{xiang:11a}, and references therein) using instruments capable
of both imaging halos and making direct measurement of metal
absorption edges (i.e., \chandra-\hetg and \xmm-\rgs), and must be
treated as having a certain degree of systematic uncertainty. However,
in regard to the values presented in Table~\ref{tab:flux}, adopting a
ratio of $\tau$ to ${\rm N_H}/10^{22}\,{\rm cm}^{-2}$ that lies
between 0.243--0.486 alters the implied X-ray fluxes by only $\pm5\%$,
our fitted ${\rm N_H}$ values by $\pm10^{22}{\rm cm^{-2}}$, and our
fitted photon indices by $\pm 0.05$.

\subsection{Results}

Our fit results are presented in Table~\ref{tab:flux} and are shown in
Figure~\ref{fig:spectra}.  Consistent with the results of the K-S test
for the single ObsID 14392, the summed quiescent spectrum is
significantly softer than the mean flare spectrum, with no overlap
between the 90\% confidence level error bars for their photon indices,
$\Gamma$. The iron emission line at 6.63 keV in the quiescent spectrum
has an equivalent width (EW) of $\approx$780 eV. If, instead of a
power-law, we fit the quiescent continuum with a thermal plasma model
that already includes ionized iron emission at a slightly higher
energy than above ({\tt vmekal}), we then find evidence for a 120\,eV
EW line at $6.3^{+0.2}_{-0.1}$\,keV, consistent with Fe\,K$\alpha$
fluorescence, and consistent with the previous suggestion of such a
line by \citet{sazonov:12a}. This model provides a similarly good fit
($\chi^{2}$/DoF$=264/256$) with plasma temperature
$kT=2.7^{+0.3}_{-0.2}$ keV and a slightly smaller X-ray absorbing
column density $N_{\rm H}=(11.8^{+0.7}_{-0.6})\times10^{22}$
cm$^{-2}$. There is good agreement with previous studies of the
quiescent spectrum (e.g., \citealt{Baganoffetal2003,sazonov:12a}),
despite the fact that these authors used larger extraction regions
(1\arcsec\hspace{-3 pt}.5 radius) and very different background
subtraction methods. We note that there is excellent agreement between
the \zth order quiescent spectrum and the \acisi spectrum.

By any measure, the bright flare mean emission observed in ObsID 14392
is significantly brighter than the quiescent emission, and assuming
that the peak spectrum has the same spectral shape as the mean
spectrum, the peak emission is a factor of 2.5 times brighter still.
The mean, absorbed 2--8\,keV flux is higher than any \sgra mean flare
flux observed with \chandra, and almost identical to the mean flux of
the brightest \sgra flare ever observed by \xmm \citep[2002
  October;][and
  Section~\ref{sec:xmm}]{Porquetetal2003,Porquetetal2008}.  With a
flare duration of $\approx 5.6$\,ks (compared to $\approx 2.8$\,ks for
the 2002 October flare), the flare's absorbed 2--8\,keV fluence is
$(4.7\pm0.5)\times10^{-8}\,{\rm erg~cm^{-2}}$, and its emitted
intrinsic energy in the 2--10\,keV band is approximately
$10^{39}$\,erg.  As is evident in Table~\ref{tab:flux}, the further we
extrapolate beyond the well-measured 2--8\,keV band, and if we
consider unabsorbed instead of absorbed fluxes, the greater the
uncertainty becomes both statistically and systematically.  

Whereas the mean, absorbed flux is extremely well-constrained, the
unabsorbed fluxes have strong dependencies upon the fitted X-ray
absorbing column.  Confidence contours for quiescent and mean
unabsorbed flare flux vs. equivalent neutral column are presented in
Figure~\ref{fig:conf}.  This figure also shows confidence contours for
fitted photon indices, $\Gamma$, vs. equivalent neutral column.
Unsurprisingly, there are strong correlations between indices and
columns, with harder photon indices being associated with lower
columns.  There are, however, two important points to note: there is a
good consistency between the fitted columns for the quiescent and
flare spectra (with the latter allowing a wider range of values owing
to the poorer statistics), and there is a very clear separation
between the index/column contours for the quiescent and flare spectra.
The flare spectrum is significantly harder; however, for most values
of X-ray absorbing column the flare photon index is somewhat softer
than implied by previous \chandra measurements of a bright flare that
indicated $\Gamma= 1\pm0.8$ (90\% confidence level;
\citealt{Baganoffetal2001}).  Absent the use of the \hetgs, however,
there is a question as to what extent this prior result was influenced
by pileup. This prior result also used an earlier version of the
\chandra calibration and did not include the $E^{-2}$ dependence of
dust scattering losses to the spectrum.

\section{Comparison to Flares Observed by \xmm}\label{sec:xmm}

Our best-fit mean flare photon index of $\Gamma = 2^{+0.7}_{-0.6}$ is
in good agreement with previous results obtained for the brightest
(2002 October 3) and second brightest (2007 April 4) \sgra flares ever
observed by \xmm \citep{Porquetetal2003,Porquetetal2008}, from 22
observations centered on \sgra which totaled 1.1\,Msec through 2009
April.  In order to carefully assess the degree to which the
properties of the \hetgs observed flare are comparable to the
brightest \xmm observed flares, we have re-fit the 2--10\,keV spectra
of the 2002 October 3 and 2007 April 4 flares --- using the same data
files as \citet{Porquetetal2008} --- with the identical
absorbed/scattered power-law model\footnote{\citet{Porquetetal2003}
  analyzed the brightest \xmm \sgra flare with a dust scattering model
  that presumed a fixed ${\rm A_V}=30$ (i.e., the scattering optical
  depth was not tied to the fitted ${\rm N_H}$) and used the
  \texttt{tbabs} absorption model and abundances of \citet{wilms:00a}
  with the cross sections of \citet{verner:96a}.
  \citet{Porquetetal2008}, in order to compare to other works, instead
  analyzed the two brightest \xmm flares with a dust scattering model
  with a fixed ${\rm A_v}=25$, and used {\tt wabs} with cross sections
  of \citet{morrison:83a} and the abundances of \citet{anders:89a}.}
presented in Section~\ref{sec:spectra}.  

For the 2002 October 3 flare we find a photon index $\Gamma =
2.3\pm0.3$, equivalent neutral column of ${\rm N_H} =
(16.1^{+1.9}_{-2.2})\times10^{22}\,{\rm cm^{-2}}$, and an unabsorbed
2--10\,keV flux of $(26.0^{+4.6}_{-3.5})\times10^{-12}\,{\rm
  erg~cm^{-2}~s^{-1}}$.  This flare's mean absorbed 2--8\,keV flux is
$(7.7\pm0.3) \times 10^{-12}\,{\rm erg~cm^{-2}~s^{-1}}$, which for its
2.8\,ks duration corresponds to an absorbed 2--8\,keV fluence of
$(2.2\pm0.1)\times10^{-8}\,{\rm erg~cm^{-2}}$ and an intrinsic emitted
energy in the 2--10\,keV band of $5.3\times10^{38}$\,erg. The photon
indices found here are very similar to those reported in
\citet{Porquetetal2008} with a $\Delta\Gamma$ of only +0.1.  The
N$_{\rm H}$ values differ due to the different abundances and
cross-sections assumed in this work.

We find a photon index of $\Gamma = 2.4^{+0.4}_{-0.3}$ for the 2007
April 4 flare, an equivalent neutral column of ${\rm
  N_H}=(16.3^{+3.0}_{-2.6})\times10^{22}\,{\rm cm}^{-2}$, and an
unabsorbed 2--10\,keV flux of
$(16.8^{+4.6}_{-3.0})\times10^{-12}\,{\rm erg~cm^{-2}~s^{-1}}$. The
mean absorbed 2--8\,keV flux is
$(4.8^{+0.2}_{-0.3})\times10^{-12}\,{\rm erg~cm^{-2}~s^{-1}}$, which
for the flare's 2.9\,ks duration corresponds to an absorbed 2--8\,keV
fluence of $(1.4\pm0.1)\times10^{-8}\,{\rm erg~cm^{-2}}$ and an
intrinsic emitted energy in the 2--10\,keV band of
$3.5\times10^{38}$\,erg.

Figure~\ref{fig:conf} also presents confidence contours of X-ray
absorbing column vs. 2--10\,keV unabsorbed flux and photon index,
$\Gamma$, for these two \xmm observed flares.  We see that the 2002
October and 2012 February events appear to be ``twin flares'' in all
respects, and that aside from having a lower flux, the 2007 April
flare otherwise appears identical to these two extremely bright
flares.  The \xmm observed flares have slightly larger best-fit values
for the equivalent neutral column; however, there is a high-degree of
overlap among the error contours.  Both \xmm observed flares are also
slightly shorter in duration, lasting $\approx 3$\,ks, with the
brightest flare lightcurve also showing a brief dip near its peak,
detected in all three \epic instruments
\citep{Porquetetal2003,Porquetetal2008}.

\section{Discussion}\label{sec:results}

Our \chandra-\hetgs observation of \sgra taken on 2012 February 9
(ObsID 14392) exhibits a flare with the highest peak flux and fluence
seen from this source. Remarkably, it is bright enough to allow the
extraction of a pure flare gratings spectrum. Our comparative analysis
indicates that in many ways, the bright \sgra flares observed by \xmm
in 2002 October and 2007 April are spectral twins to this
flare. Depending upon how one defines and measures \sgra's quiescent
flux and in how one extrapolates the mean flare flux to a peak flux,
for both the 2012 February and 2002 October flares, the ratio of peak
flux to quiescent flux is at least a factor of order 130.  If one
adopts the suggestion of \citet{shcerbakov:10a} that only $\approx
1\%$ of the observed quiescent flux is from the central regions (where
the flare likely originates), then this factor is more plausibly of
order $10^4$!

Given systematic uncertainties in extrapolating unabsorbed fluxes and
defining the quiescent flux associated solely with the \sgra point
source, however, we suggest that a less ambiguous set of reported
values are the flare's mean absorbed 2--8\,keV flux, its absorbed
2--8\,keV fluence, and the ratio of its peak rate to mean rate, each
measured as values \emph{above} the quiescent level and aperture
corrected for the instrument's PSF.  So long as instrumental spectral
extraction regions are consistent for quiescent and flare spectra,
these values will be well-defined.  (However, due to the short time
scale sub-structure in the flare, as seen in Figure~\ref{fig:BB} and
previously reported by \citealt{Porquetetal2003}, the ``peak'' flux
value may actually be difficult to define precisely.)  The 2--8\,keV
band is above the range of the most severe X-ray absorption in \sgra
and is well-covered by the three soft X-ray instruments best capable
of observing \sgra: \chandra, \xmm, and \swift.

The bright flare is asymmetric, with a slower rise than decay. This
may be due to unresolved sub-structure: the Bayesian Blocks
decomposition shows evidence of complex structure near the flare peak,
and a fit with two Gaussian profiles, with different widths and offset
from one another, works well.  (The fitted Gumbell profile is a
convenience that allows us to easily calculate the flare peak/mean
ratio.)  Alternatively, if this is the flare's intrinsic profile it is
quite different than the ``fast rise, exponential decay'' of many
different types of transient phenomena.

The total emitted 2--10\,keV energy of the flare, which is of ${\cal
  O}(10^{39}$\,erg$)$, requires conversion of at least $10^{19}$\,g of
rest mass into energy, presuming a 10\% conversion efficiency. There
is no agreed upon mechanism for flare energization, with suggestions
having ranged from magnetic reconnection
\citep{Markoffetal2001,YuanQuataertNarayan2003,Dodds-Edenetal2010} to
tidal disruption of asteroids
\citep{cadez:08a,kostic:09a,ZuvobasNayakshinMarkoff2012}.  Regardless
of the energization mechanism, with the results from the prior \xmm
observations it appears that at least these extremely bright flares
require an emission mechanism that produces a moderate (i.e., not very
hard) photon index, $\Gamma \approx 2$.

As we have shown, much of the uncertainty on the spectral slope is
systematic, and depends on assumptions made about absorption and
scattering.  Reasonable assumptions about the scaling between dust
scattering and absorption have only a small systematic effect ($\Delta
\Gamma \sim 0.05$), but the dependence upon fitted ${\rm N_H}$ is more
pronounced (i.e., Figure~\ref{fig:conf}).  Although there is a good
consistency between the equivalent neutral column for the quiescent
and flare spectra, it is possible to alter the fitted ${\rm N_H}$ by
assuming different metal abundances, or metal depletions in dust
grains, or different cross sections, etc. Based on the correlation
from \citet{predehl:95a} and our assumptions about $N_{\rm H}$ and
${\rm A_V}$ outlined in Section~\ref{sec:spectra}, the minimum
reasonable X-ray absorbing column for the quiescent spectrum implies an
extinction of ${\rm A_V}\approx 40$. This is somewhat at odds with
prior estimates of $\rm A_{V}$: a discussion of the Galactic center
optical and infrared extinction curves can be found in
\citet{fritz:11a}, who suggest a value of ${\rm A_V} \approx 33$ (in
part based upon the X-ray observations of \sgra discussed in
\citealt{Porquetetal2003}). The relationships derived by
\citet{watson:11a} and \citet{guver:09a}, ${\rm A_V} \approx N_{\rm
  H}/2.2\times10^{21}\,{\rm cm^{-2}}$, imply a similar value for the
extinction towards \sgra. However, the methods used to derive $N_{\rm
  H}$ and ${\rm A_V}$ in these works are very heterogeneous, and it is
unclear if their scalings can be extrapolated to $\rm A_V\sim40$. It
is therefore imperative that scaling between $N_{\rm H},$ extinction,
and dust scattering optical depth be revisited with modern models
using consistent cross sections and abundances, and with modern high
spectral and imaging resolution observations. For the moment, we
conclude that the extinction, dust scattering optical depth, $N_{\rm
  H}$, and (by extension) the spectral properties of \sgra are still
subject to a certain degree of systematic uncertainty.

Despite these uncertainties, we find similar spectral properties for
the brightest \chandra and \xmm flares ($\Gamma\sim$2 and $N_{\rm
  H}\sim15\times10^{22}$ cm$^{-2}$).  Weaker flares could in principle
have harder spectra than bright flares, but their spectral properties
are not yet yet strongly constrained (as shown and discussed in
\citealt{Porquetetal2008}).  However, there is significant cause for
optimism, as the observations discussed in this work represent
slightly more than 10\% of the \chandra-\hetgs observations of \sgra
that will occur in 2012. We anticipate that this program will detect
over three dozen flares, with perhaps one or two more with amplitudes
comparable to or greater than that of the 2012 February flare. As we
have demonstrated that we can extract \chandra gratings spectra for
\textit{individual} bright flares, the XVP program should provide
unprecedented constraints on the spectral properties of faint and
moderate flares, which will allow us to determine how the physics of
flares scales with their luminosity.

For the first time, the spatially resolved \chandra studies will
produce \sgra flare spectra that are either absent of detector pileup
or that have pileup strongly constrained (as is the case here) owing
to the simultaneous \fst order gratings spectra. Our understanding of
the physics of flares will be greatly enhanced by the fact that many
of the upcoming observations will be performed with simultaneous
multi-wavelength observations. The \chandra XVP program will offer us
the unique opportunity to study the physics underlying accretion onto
\sgra and other quiet galaxies.

\acknowledgements We would like to thank all the members of the \sgra
\chandra-XVP
team\footnote{\texttt{www.sgra-star.com/index.php/menu-collaboration-members}}.
We would also like to thank the \chandra schedulers for their
continued superb support of this challenging program.  We thank J\"orn
Wilms for useful discussions concerning absorption and extinction
models. Michael Nowak, Joey Neilsen, and John Houck gratefully
acknowledge funding support from the National Aeronautics and Space
Administration through the Smithsonian Astrophysical Observatory
contract SV3-73016 to MIT for support of the \chandra X-ray Center,
which is operated by the Smithsonian Astrophysical Observatory for and
on behalf of the National Aeronautics Space Administration under
contract NAS8-03060. Fred Baganoff was supported by NASA Grant
NAS8-00128.  Fred Baganoff, Michael Nowak, and John Houck also were
supported by NASA Grant GO2-13110A.  Sera Markoff is grateful for
support from a Netherlands Organization for Scientific Research (NWO)
Vidi Fellowship, as well as from The European Community's Seventh
Framework Programme (FP7/2007-2013) under grant agreement number ITN
215212 ``Black Hole Universe''. This work was based in part on
observations obtained with \xmm, an ESA science mission with
instruments and contributions directly funded by ESA Member States and
NASA.


\end{document}